\documentclass{aa}
\usepackage{graphicx}
\usepackage{txfonts}
\usepackage{epsfig}
\def\ltsima{$\; \buildrel < \over \sim \;$}
\def\gtsima{$\; \buildrel > \over \sim \;$}
\def\simlt{\lower.5ex\hbox{\ltsima}}
\def\simgt{\lower.5ex\hbox{\gtsima}}

\begin{document}

   \title{XMM-$Newton$ observations of 4 luminous radio-quiet AGN, and the soft X--ray excess problem}

   \author{F. D'Ammando \inst{1,2},  S. Bianchi\inst{1}, E. Jim\'enez-Bail\'on\inst{1,3,4}, G. Matt\inst{1} }

   \offprints{F. D'Ammando, \email{filippo.dammando@iasf-roma.inaf.it} }

   \institute {$^1$Dipartimento di Fisica, Universit\`a degli Studi Roma Tre, via della Vasca Navale 84, I-00146 Roma, Italy \\
$^2$Dipartimento di Fisica, Universit\`a di Roma Tor Vergata, via della
   Ricerca Scientifica 1, I-00133 Roma, Italy \\
$^3$ LAEFF-INTA, Apdo. 50727, 28080-Madrid, Spain \\
$^4$ Universidad Nacional Autonoma de Mexico, Apartado Postal 70-264, 04510 Mexico, DF, Mexico
}

   \date{Received / Accepted }

   \abstract{The nature and origin of the soft X--ray excess in radio--quiet AGN is still an open issue. The
interpretation in terms of thermal disc emission has been challanged by the discovery of the constancy of the
effective temperature despite the wide range of Black Hole masses of the observed sources.
Alternative models are reflection from ionized matter and absorption in a relativistically
smeared wind.}
{ We analyzed XMM--$Newton$ observations of four luminous radio--quiet AGN with the aim
of characterising their main properties and in particular the soft excess.}
{Different spectral models for the soft excess were tried: thermal disc emission, Comptonization,
ionized reflection, relativistically smeared winds.}{Comptonization of thermal emission and the smeared winds provide
the best fits, but the other models also provide acceptable fits. All models, however, return
parameters very similar from source to source, despite the large differences in luminosities, Black Hole
masses and Eddington ratios. Moreover, the smeared wind model require very large smearing velocities.
The UV to X--ray fluxes ratios are very different, but do not correlate with any other parameter.}{No fully
satisfactory explanation for the soft X--ray excess is found. Better data, like e.g. observations in a broader
energy band, are needed to make further progresses. }

\authorrunning{F. D'Ammando et ~al. }
\titlerunning{XMM-$Newton$ observations of 4 luminous AGN}

   \maketitle
%

\section{Introduction}

While relatively low luminosity AGN are well studied in X-rays, not the same can be said 
for high luminosity ones, due to their paucity in the local Universe and therefore on  
their relative low fluxes. With the aim to populate the high L -- high flux
portion of the parameter space, we proposed and obtained XMM--$Newton$ observations of four sources, 
selected from the Grossan HEAO1 LMA catalogue to be bright 
(F$_{2-10}$ $>$ 10$^{-11}$ erg cm$^{-2}$ s$^{-1}$) and luminous (L$_{2-10}$ $>$ 10$^{44}$ erg s$^{-1}$): 
H0439-272, Ark 374, Fairall 1116 and PG0052+251 (see Table 1). 

HEAO1 instruments scanned the whole sky from 1977 to 1979. 
The A-1 instrument Large Sky Survey (LASS, Wood et al. 1984) catalogue contains 322 sources 
brighter than 0.0036 LASS counts/s/cm$^2$ (F$_{2-10}$$>$ 1.82 $\cdot$ 10$^{-11}$ erg cm$^{-2}$ s$^{-1}$, 
assuming a power law spectrum with $\Gamma$ = 1.7) and with $\mid$b$\mid$ $>$ 20 
(Grossan, 1992).
The A-1 instrument consisted of a set of proportional counters sensitive in the 2-20 keV band, 
which observed the sky through passive collimators with field of view of 1x4 and 0.5x1 degrees.
The precision with which the position of each X-ray source was determinated was greatly improved by 
the simultaneous observation of the MC instrument (Modulation Collimator or A-3). The MC instrument 
produced a pattern of small, diamond-shaped error regions, only 1x4 arcmin each (90$\%$ 
confidence limit).
The size of the final LASS/MC error box is therefore of $\sim$ 0.3 deg$^2$. 
Grossan and collaborators identified optical counterparts of 287 of the 322 objects in the LASS 
catalog (86$\%$). 96 of these objects are identified with AGN: the LASS-MC-AGN (LMA) sample 
(Grossan, 1992), by correlation with catalogs of known sources and through optical photometry 
and spectroscopy performed by the MIT group.

Information on the four sources and on the related observations are summarized in Tables~\ref{log} 
and ~\ref{fluxes}.
Fluxes are derived from the best fit models (see Sec.3). Unfortunately, the sources were 
all found at fluxes (and therefore luminosities) lower by at least a factor 2 than in the
HEAO-1 observations (not unexpectedly, as they all were just above the flux threshold). 
Fortunately, they were still bright enough to allow for a detailed spectral analysis. 

In this paper, we report on the spectral analysis of these four sources, with particular
emphasis on the soft excess.

A soft excess is present in the X-ray spectra of most Seyfert 1s and radio-quiet quasars. 
First noted by Arnaud et al. (1985), it is an emission below $\sim$ 1 keV in excess of the 
extrapolation of the power law component dominating at higher energies. 

The origin of the soft excess is still an open issue. In the past, it was often associated with the
thermal emission of the accretion disc. The large effective temperatures usually obtained (when compared to 
what expected in a standard Shakura $\&$ Sunyaev, 1973, accretion disc), were attributed
to a slim accretion disc in which the temperature is raised by photon trapping 
(Abramowicz et al. 1998, Mineshige et al. 2000), in which case the accretion is super-Eddington, 
or to Comptonisation of EUV accretion disc photons (Porquet et al. 2004).

However, it has been recently shown that the disc
temperature is remarkably constant around 0.1 - 0.2 keV, 
regardless of the central object luminosity and mass 
(Gierlinski $\&$ Done 2004, Crummy et al. 2006). This result is 
difficult to explain in any model for the soft excess related to
disc continuum emission, as in any disc model the temperature is 
expected to vary with both the Black hole mass and the accretion rate.

On the other hand, the constancy of the temperature may  
arise naturally if the soft excess is not related to thermal
emission at all, but due to absorption/emission processes, 
the ``temperature'' actually measuring atomic transitions.
In fact, there is a great increase in opacity in partially ionized material 
due to lines and edges of ionized O VII, O VIII and Iron at $\sim$ 0.7 keV, 
which can make the apparent soft excess, provided that 
velocity smearing broadens and blurs the atomic features.

On this line of tought, two alternative scenarios have been proposed:
a partially ionized, relativistically blurred reflection from the accretion disc 
(Crummy et al. 2006) and a velocity smeared, partially ionized absorption 
(Gierlinski $\&$ Done 2004), the latter model further developed by Schurch $\&$ Done (2006) 
by including the emission associated with the absorbing material 
and proposing an origin in a `failed wind'. 

In the reflection model, the partially ionized material is 
naturally identified with the accretion disc, but to produce a strong soft 
excess the intrinsic continuum must be significantly 
suppressed, perhaps by disc fragmentation or by 
light--bending effects (Fabian et al. 2002, 2004, 2005; Miniutti $\&$ Fabian 2004).

In the absorption model, the partially ionized material is supposed to be a
wind likely starting from the disc (Schurch $\&$ Done 2006). If this is the case,
the velocity structure  has to be complex, giving rise to a substantial broadening 
to mask the sharp atomic features, which otherwise should have been observed in high
resolution gratings observations.  
A problem with this interpretation, as noted by the same authors, is that such winds tend to 
be produced within $\sim$ 25$^{\circ}$ of the equatorial plane. 
Most of type-1 AGN are expected to have lines of sight which would not intercept much of this material, 
but this is in conflict with the fact that most type-1 AGN show a soft excess.\ 
Moreover, the required large velocity smearing implies a mass-loss rate 
much larger than the accretion rate required to power the observed luminosity. Schurch $\&$ Done (2006)
then proposed a `failed wind' instead, which does not require a large mass-loss rate.
In such a case, the central X-ray source is strong enough to overionize the wind removing 
the acceleration before the material reaches escape velocity, allowing the material to fall back to the disk.

\section{Observations and data reduction}

\begin{table}
\begin{center}
\begin{tabular}{cccccc}
\textbf{Source} & \textbf{Date} &\multicolumn{2}{c}{\textbf{EPIC-pn}}& \textbf{z} & $\mathbf{N_\mathrm{Hg}}$ \\
& & \textbf{T (ks)} & & & $\mathbf{10^{20}}$ \textbf{cm}$\mathbf{^{-2}}$ \\
\hline \textbf{H0439-272} & 2005-08-13 & 20 & & 0.084 & 2.50\\
\hline \textbf{Ark 374} & 2005-07-09 & 25 & & 0.063 & 2.21\\
\hline \textbf{Fairall 1116} & 2005-08-28 & 20 & & 0.058 & 3.29\\
\hline \textbf{PG 0052+251} & 2005-06-26 & 20 & & 0.155 & 4.81 \\
\hline
\end{tabular}
\caption{For each source: dates and exposure times of the observation,  redshift and line--of--sight 
Galactic absorption.}
\label{log}
\end{center}
\end{table}

\begin{table*}
\begin{center}
\begin{tabular}{cccccccc}
\textbf{Source} & \textbf{$F_{0.5-2}$} & \textbf{$F_{2-10}$} & \textbf{$L_{2-10}$} & \textbf{$L_B$} & \textbf{$M_{BH}$} & $\lambda$ & FWHM $H_{\beta}$ \\
& ($10^{-12}$ erg cm$^{-2}$ s$^{-1}$) & ($10^{-12}$ erg cm$^{-2}$ s$^{-1}$) & ($10^{43}$ erg s$^{-1}$) & ($10^{43}$ erg s$^{-1}$) & (10$^8$ $M_{\odot}$) &  & km/s \\
\hline \textbf{H0439-272} & 4.54 & 5.74 & 10.0 & 34.0 & 0.36  & 0.73 & 2500 \\
\hline \textbf{Ark 374} & 3.16 & 3.29 & 3.1 & 7.2 & 0.74-1.4  & 0.04-0.075 & 4200 \\
\hline \textbf{Fairall 1116} & 4.58 & 5.42 & 4.3 & 11.1 & 1.3  & 0.065 & 4310\\
\hline \textbf{PG 0052+251} & 5.03 & 6.77 & 42.6 & 230 & 3.7-8.5  & 0.21-0.48 & 4165 \\
\hline
\end{tabular}
\caption{For each source: 0.5--2 and 2-10 keV flux  and luminosity; 
bolometric luminosity (derived adopting Marconi et al. (2004) bolometric corrections); 
Black Hole mass, Eddington ratio and $H_{\beta}$ FWHM. Black Hole masses are from: Shields et al. (2003) 
for Ark~374 (lower  value) and Fairall~1116; from Vestergaard \& Peterson (2006) for Ark~374 
(upper value) and PG~0052+251; and from the H$\beta$ FWHM reported in Grupe et al.
(2004) for H0439-272, using the formula in Vestergaard \& Peterson (2006). Other values of the 
H$\beta$ FWHM are from
Corbin 1991 (Ark~374), Grupe et al. 2004 (Fairall~1116) and Peterson et al. 2004 (PG~0052+251).} 
\label{fluxes}
\end{center}
\end{table*}

All the sources were observed by XMM-$Newton$ with all the EPIC CCD cameras, the p-n (Str\"uder et
al. 2001) and the two MOS (Turner et al. 2001) cameras, with the RGS (der Herder et al. 2001) and the OM
(Mason et al. 2001). 
To reduce pile-up to negligible values (about 0.1$\%$ in the p-n) we decided to adopt small window 
modes for all sources and all EPIC instruments, but for MOS1, for which we used a full window mode 
in order to check for possible confusion problems within the HEAO1 error box.

Data were reduced with the Science Analysis System (SAS) v.7.1.0 software package, 
adopting standard procedures, while screening 
for intervals of flaring particle background in the EPIC data was done consistently with the choice of the 
extraction radii, in order to maximise the signal-to-noise ratio, similarly to what described by 
Piconcelli et al. (2004).

We do not make use of MOS data in this paper because the MOS are slightly piled-up, despite the small 
window mode adopted for MOS2, and the addition of the MOS in the spectral fits do not significantly 
increase the precision with which spectral parameters are determinated. We do not make use of the RGS
data either, because an inspection of their spectra shows they are featureless. 

P-n spectra were rebinned in order to have at least 20 counts/bin and to oversample the energy resolution 
by a factor about 3. Patterns 0 to 4 were included in the p-n spectrum, whose count rate is lower than the 
maximum for 1$\%$ pile-up (see Table 3 of Ehle et al. 2005). 

Five exposures with the Optical Monitor were also available, all with the UVM2 (231 nm) filter. 
The Observation Data Files (ODF) of the OM for each source were extracted and processed using 
the meta-task $omichain$ of the XMM $SAS$, with standard procedures.

All spectra were analyzed using Xspec v.12.3.1 (Arnaud 1996). In the following, all energies are in the 
source rest frame, and errors corresponding to 90\% confidence level for one interesting parameter 
($\Delta\chi^2$=2.71), unless otherwise stated.

Luminosities were calcolated considering a flat $\Lambda$CDM cosmology with ($\Omega_M$, $\Omega_\Lambda$) = 
(0.3, 0.7) and H$_0$ = 70 Km s$^{-1}$ Mpc$^{-1}$ (Bennett et al. 2003).
Metal abundances have been kept fixed to cosmic ones, according to Anders \& Grevesse (1989).

\section{Data Analysis}

\subsection{Baseline fit model}

\begin{figure*}
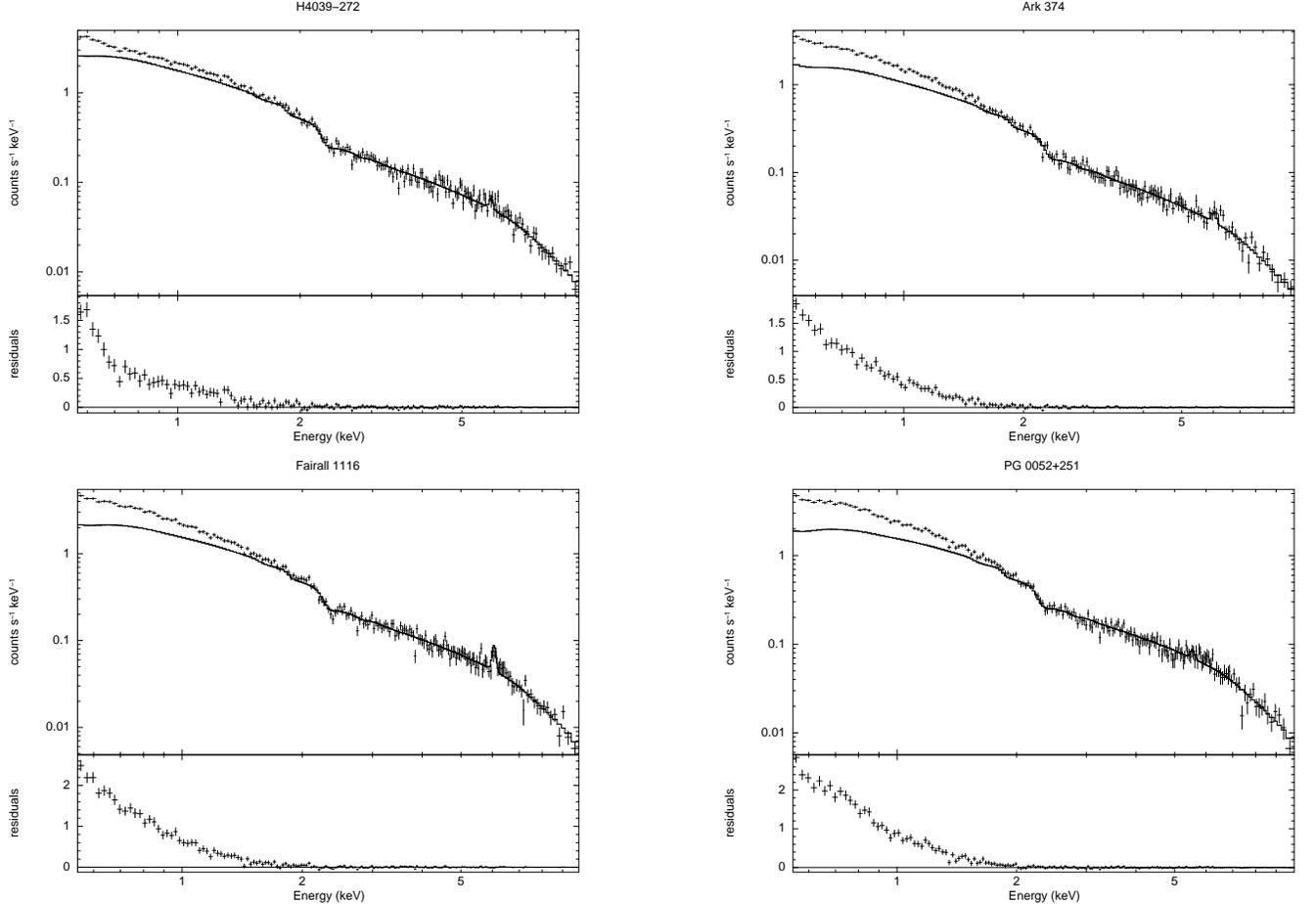

\begin{minipage}[tp!]{0.525\textwidth}
\begin{center}
\epsfig{file=H0439_soft.ps,width=6cm,angle=-90}
\end{center}
\end{minipage}
\medskip
\begin{minipage}[tp!]{0.525\textwidth}
\begin{center}
\epsfig{file=Ark374_soft.ps,width=6cm,angle=-90}
\end{center}
\end{minipage}
\medskip
\begin{minipage}[tp!]{0.525\textwidth}
\begin{center}
\epsfig{file=Fairall1116_soft.ps,width=6cm,angle=-90}
\end{center}
\end{minipage}
\medskip
\begin{minipage}[tp!]{0.525\textwidth}
\begin{center}
\epsfig{file=PG0052_soft.ps,width=6cm,angle=-90}
\end{center}
\end{minipage}
\caption{Spectrum and best fit model (upper panel) and residuals (lower panel) of the 4 sources, 
when fitted with a single power law in the 2-10 keV
band, and then extrapolated to lower energies.}
\label{base}
\end{figure*}

We started the analysis by fitting the p-n instrument only, in the 0.5-10 keV energy range.
A simple power law (absorbed by the Galactic column) gives unacceptable fits, with strong
excess residuals at low energies (see Fig.~\ref{base}). 
The inclusion of a neutral Compton reflection component
({\sc pexrav} in Xspec) does not help much, unless very large and unrealistic (given also
the iron line equivalent widths, see below) values for the relative normalizations 
are allowed. We then added a further power law
to reproduce the soft excess. Let us call it the  baseline model spectrum (BMS), which in summary 
consists of: two power laws, cold absorption from the Galactic hydrogen column density; a Gaussian line to 
reproduce the iron K$\alpha$ fluorescent emission line. We fixed the rest frame energy of the line to 6.4 keV 
and the width at 10$^{-3}$ keV; the fits did not improve when these parameters were left free to vary (see below).

The fit of the BMS was very good for all sources  (reduced $\chi^{2}$ about 1), except for H0439-272,
for which inspection of residuals suggested the inclusion of an absorption edge at $\sim$0.72 keV
(consistent with the O {\sc vii} K edge) which indeed reduces
the $\chi^2$ from 251 (186 d.o.f.) to 204 (184 d.o.f.).
The best fit parameters for BMS (here and after including the absorption edge for H0439-272)
are summarized in table~\ref{fitnosoft}.

\subsection{The iron line}

Two sources, Ark 374 and Fairall 1116, have a strong enough iron line to search for complexities in their profiles.

For Ark 374, leaving the energy and width of the line free to vary, a marginal 
improvement is found ($\Delta\chi^2$=5.6). The best fit centroid energy and width are 6.48$^{+0.20}_{-0.07}$ keV
and 0.18$^{+0.14}_{-0.08}$ keV, respectively. Even less significant is the improvement for Fairall 1116 ($\Delta\chi^2$=1.4).
The best fit centroid energy and width are 6.43$^{+0.03}_{-0.04}$ keV
and $<$0.09 keV, respectively. For all sources, therefore, we can conclude that the iron line is consistent
with being neutral and unresolved, even if a relativistic disc component cannot, of course, be excluded given
the modest quality of the data. 

It is interesting to note that, even considering the large error bars, 
the Equivalent Width of the iron line seems to diminishes with the 
luminosity (the Iwasawa--Taniguchi effect) and with the Eddington ratio, 
as found by Bianchi et al. (2007) in a much larger sample.

\begin{table*}
\begin{large}
\vspace{1.5mm}
\centering
\begin{tabular}{c c c c c c}
\textbf{Object} & $\Gamma_{1}$ & $\Gamma_{2}$ & EW$_\mathrm{Fe}$&$\chi^{2}_r$/d.o.f. & Null Hyp. Prob.\\
  & & & (eV) & & \\
\hline \textbf{H0439-272}   & $1.54^{+0.15}_{-0.37}$ & $2.63^{+0.33}_{-0.23}$ &  54$\pm$36  &  1.11/184 & 0.15\\
\textbf{Ark 374}   &$1.43^{+0.26}_{-0.23}$ & $2.72^{+0.22}_{-0.16}$ & 71$^{+50}_{-42}$  & 0.97/167 & 0.61\\
\textbf{Fairall 1116}  & $1.56^{+0.10}_{-0.23}$ & $2.93^{+0.17}_{-0.23}$ & 113$\pm$41 & 1.01/190 & 0.44\\
\textbf{PG0052+251}  & $1.25^{+0.12}_{-0.20}$ & $2.71^{+0.14}_{-0.12}$ & $<$61 & 0.92/192 & 0.77\\
\hline
\end{tabular}
\caption{\label{fitnosoft}Best fit parameters for the BMS: wabs (powerlaw+powerlaw+zgauss).
 In the case of H0439-272, an edge
with $E_{\rm th}$=0.72$\pm$0.02 and $\tau$=0.27$\pm$0.05 has been added. $^*$fixed}
\end{large}

\begin{large}
\vspace{1.5mm}
\centering
\begin{tabular}{c c c c c}
\textbf{Object} &\textbf{kT}(keV) & $\Gamma$ & $\chi^{2}_r$/d.o.f. & Null Hyp. Prob.\\
\hline \textbf{H0439-272}  & $0.21^{+0.02}_{-0.02}$ & $1.90^{+0.04}_{-0.04}$ & 1.14/184 & 0.09 \\
 \textbf{Ark 374}  & $0.18^{+0.01}_{-0.01}$ & $1.97^{+0.04}_{-0.05}$ & 1.15/167 & 0.09  \\
 \textbf{Fairall 1116} & $0.18^{+0.01}_{-0.01}$ & $1.89^{+0.04}_{-0.04}$ & 1.14/190 & 0.09 \\
 \textbf{PG0052+251} & $0.19^{+0.01}_{-0.01}$ & $1.81^{+0.04}_{-0.04}$ & 1.12/192 & 0.11 \\
\hline
\end{tabular}
\caption{Best fit parameters for the model: wabs ( diskbb + powerlaw + zgauss ).}\label{diskbb}
\end{large}

\begin{large}
\vspace{1.5mm}
\centering
\begin{tabular}{c c c c c c}
\textbf{Object} &\textbf{kT}(keV) & $\tau$ & $\Gamma$ & $\chi^{2}_r$/d.o.f. & Null Hyp. Prob.\\
\hline 
\textbf{H0439-272}  & $0.46^{+5.74}_{-0.18}$ & 20$^{+8}_{-18}$ & $1.79^{+0.08}_{-0.18}$ & 1.10/185 & 0.16 \\
 \textbf{Ark 374}  & $0.41^{+3.30}_{-0.11}$ & 21$^{+6}_{-6}$ & $1.80^{+0.08}_{-0.16}$ & 0.96/166  & 0.62  \\
 \textbf{Fairall 1116} & $0.38^{+0.32}_{-0.10}$ & 22$^{+6}_{-5}$ & $1.76^{+0.08}_{-0.08}$ & 1.00/189  & 0.49\\
\textbf{PG0052+251}  & $0.39^{+0.26}_{-0.07}$ & 22$^{+3}_{-6}$ & $1.63^{+0.05}_{-0.10}$ & 0.91/191 & 0.81 \\
\hline
\end{tabular}
\caption{Best fit parameters for the model: wabs ( compst + powerlaw + zgauss ). }
\label{compst}
\end{large}

\begin{large}
\vspace{1.5mm}
\centering
\begin{tabular}{c c c c c c c }
\textbf{Object} & $\xi$ & $\Gamma$ & Fraction & A$_{Fe}$  & $\chi^{2}_r$/d.o.f. & Null Hyp. Prob.\\
\hline
\hline \textbf{H0439-272}  & $465^{+120}_{-85}$ & $1.96^{+0.01}_{-0.02}$ & 0.14  & $1.24^{+0.48}_{-0.35}$ & 1.15/185  & 0.07 \\
 \textbf{Ark 374}  & $550^{+103}_{-174}$ & $2.03^{+0.04}_{-0.03}$ & 0.22  & $1.01^{+0.31}_{-0.16}$ & 1.23/166 & 0.02\\
 \textbf{Fairall 1116}  & $441^{+76}_{-49}$ & $1.94^{+0.03}_{-0.03}$ & 0.25  & $1.07^{+0.23}_{-0.15}$ & 1.15/189 & 0.08 \\
 \textbf{PG0052+251}  & $445^{+75}_{-30}$ & $1.90^{+0.02}_{-0.03}$ & 0.27 & $1.34^{+0.34}_{-0.27}$ & 1.14/191 & 0.09\\
\hline \textbf{H0439-272}  & $585^{+340}_{-160}$ & $1.96^{+0.03}_{-0.03}$ & 0.17  & $0.97^{+0.64}_{-0.27}$ &   1.15/185  & 0.07 \\
 \textbf{Ark 374}  & $728^{+590}_{-193}$ & $2.02^{+0.03}_{-0.04}$ & 0.25  & $0.86^{+0.25}_{-0.19}$ & 1.10/166 & 0.18\\
 \textbf{Fairall 1116}  & $560^{+130}_{-110}$ & $1.93^{+0.03}_{-0.03}$ & 0.30  & $0.94^{+0.26}_{-0.18}$ & 1.05/189 & 0.32 \\
 \textbf{PG0052+251}  & $530^{+84}_{-72}$ & $1.91^{+0.03}_{-0.03}$ & 0.36 & $0.92^{+0.34}_{-0.18}$ & 1.16/191 & 0.06\\
\hline
\hline
\end{tabular}
\caption{Best fit parameters for the model: wabs ( kdblur(reflion) + powerlaw + zgauss ). The upper 4 rows
refer to the emissivity index fixed to 2, the lower to 3. See text for further details.}
\label{reflecti}
\end{large}

\begin{large}
\vspace{1.5mm}
\centering
\begin{tabular}{c c c c c c c }
\textbf{Object} & Column & log $\xi$ & $\sigma$ & $\Gamma$ & $\chi^{2}_r$/d.o.f. & Null Hyp. Prob.\\
& $\times10^{22}$ cm$^{-2}$ & & (v/c) &  & &\\
\hline \textbf{H0439-272}  & $28.8^{+21.2*}_{-12.3}$ & $3.41^{+0.19}_{-0.17}$ & $0.5^*$ & $2.04^{+0.03}_{-0.04}$ & 1.11/184 & 0.15 \\
 \textbf{Ark 374}  & $27.3^{+11.0}_{-9.2}$ & $3.32^{+0.14}_{-0.12}$ & $0.5^*$ & $2.13^{+0.04}_{-0.03}$ & 1.03/167 & 0.38\\
 \textbf{Fairall 1116}  & $21.7^{+7.7}_{-6.5}$ & $3.21^{+0.11}_{-0.15}$ & $0.46^{+0.04*}_{-0.10}$ & $2.09^{+0.03}_{-0.03}$ & 1.02/189 & 0.39\\
 \textbf{PG0052+251}  & $35.4^{+14.6*}_{-8.4}$ & $3.35^{+0.10}_{-0.12}$ & $0.46^{+0.04*}_{-0.12}$  & $2.04^{+0.03}_{-0.04}$ & 0.94/191 & 0.72\\
\hline
\end{tabular}
\caption{Best fit parameters for model: wabs ( swind1 (powerlaw) + zgauss ) $^*$ parameter fixed, or 
error pegged to the boundary (see text for details).
} 
\label{swind}
\end{large}
\end{table*}

\subsection{The Soft Excess}

For all sources, a soft X-ray emission in excess of the extrapolation of the hard power 
law is clearly present (see Fig. 1).
In the BMS model, the soft excess was reproduced by a simple power law.  
To investigate the nature of this component, we then tried alternative models, 
representative of the different possible origins (for H0439-272 we also included
the absorption edge as detailed in Sec.3.1).

An important characteristic of the soft excess observed in previous studies is the constancy of its 
temperature when fitted with a thermal component (e.g. Gierlinki \& Done 2004), contrary to the expected
$\propto M_{BH}^{-{1 \over 4}}$ relation, suggesting that the excess is related to atomic processes.

Firstly, we checked whether the temperature is constant also in our small sample by substituting the
soft power law with an accretion disc model consisting of multiple blackbody components 
(the DISKBB model in Xspec) to reproduce the soft excess (table~\ref{diskbb}). The fits 
are acceptable but worse than with the 2 power laws model. The temperature at the inner disc radius
is very high and
similar for all the sources (which have Black Hole masses different by up to an order of magnitude).
The values are similar to those found by e.g. Piconcelli et al. (2005).

Then we tried to reproduce the soft excess with a second Comptonization region (table~\ref{compst}), using 
the simplest model available in Xspec, i.e. {\sc compst} (Sunyaev $\&$ Titarchuk 1980). This second region could
be identified with a hot disc surface layer where electrons upscatter the photons emitted in deeper layers.
The temperature of the electrons $kT_e$ and the optical depth $\tau$ are again, and surprisingly, very much the same
for all the sources. (The energy of the injected photons is hidden in the normalization parameter; in any case
the spectral shape is determined by the electron temperature and the optical depth of the hot layer). 
The quality of the fits are similar to those with the two power laws, and better than those with
the multicolor disc model.

We then tried the two models which explain the soft excess in terms of atomic physics processes, namely the
ionized disc reflection and the smeared absorption.
To describe the reflection from the accretion disc we used the latest publicly available version of the 
code described by Ballantyne, Iwasawa $\&$ Fabian (2001; see also Ross \& Fabian 2005 and Crummy et al. 2006), 
corrected by a convolution model for relativistic smearing ({\sc kdblur} in Xspec).
The reflection is characterized by the ionisation parameter of the reflecting matter, $\xi$, the photon index of 
the incident power law (which we assumed to be the same as the observed power law)
and the reflection parameter. The parameters which characterize the relativistic smearing 
are the spin of the Black Hole, the inner and outer radii of the reflecting disc, 
the emissivity (parametrized as a power law) and the inclination angle. In practice, and for the sake of
simplicity, we assumed a maximally rotating Black Hole ($a$=0.998) and 
fixed the emissivity index to 2, the inclination angle to 30$^{\circ}$, the inner and outer radii
to 1.23 and 400 gravitational radii. The results are summarizeded in table~\ref{reflecti}. 
In the table, the parameter ``Fraction'' indicates
the ratio of the reflected to direct continuum at 15 keV, where the effects of ionization are small and the ratio
is dominated by the geometry. For comparison, a neutral reflection component from a face-on slab
with R=1, $\Gamma$=2 and $A_{Fe}$=1 gives a fraction of 0.44. The quality of the fit are worse than with the
previous model. This may be (at least partly) due to our choice of keeping the disc parameter fixed (but leaving
the inner radius and the inclination angle free to vary, a significantly better fit, i.e. $\chi^2_r$=0.96/190 d.o.f.,
is found only for PG0052+251, due to a much higher inclination,  86$^{\circ}$ ) and the 
oversimplifications of the model like, e.g. a single ionization zone and a simple power law model for the
emissivity law, which instead is much more complicated (see e.g. Martocchia et al. 2002) for the lamp-post geometry. 
The fraction of the reflection component ranges from 0.14 to 0.27, i.e. between 1/3 and almost 2/3 of a whole
disc seen face--on.  The iron abundance is always about solar, while the ionization parameter spans a relatively
small range of values. If the emissivity law is set to 3, the quality of the fit is similar in H0439-272 and
PG0052+251, and better (but still worse than with previous models) in Ark~374 and Fairall~1116 (see Table \ref{reflecti}). 
The best fit parameters are all consistent within the errors with those obtained with an emissivity index of 2, even if the 
values of the ionization parameter are always larger and the iron abundance smaller.

Finally, we replaced the reflector with a relativistically smeared absorption 
(the SWIND1 model, developed by Gierlinski $\&$ Done 2006; see table~\ref{swind}), 
even if this model has been recently criticized by Schurch \& Done (2007a,b) as being oversimplified.
The SWIND1 model has 3 parameters: the ionization, the density column and the Gaussian velocity smearing, $\sigma$.
The fits are as good as the ones with two power laws. The column density and the ionization parameter
of the absorbing wind are very similar one another, and the smearing terminal velocity very high, close to 0.5 c.
Fixing the latter parameter to a smaller value, 0.15 c, resulted in much worse, unacceptable fits 
($\chi_{r}^{2} >$1.38, the latter value being obtained for the faintest source, Ark 374). 
Note that when a refined version of the model (not yet available for spectral fitting) 
 is considered, Schurch \& Done (2007b) found that indeed 
extremely large, unrealistic smearing terminal velocities are required.

\subsection{The Ultraviolet flux}

The UV fluxes of the Optical Monitor observations were obtained by simply
converting the count rates with the method 1 described in 
the SAS documentation.\footnote{\scriptsize\tt
http://xmm.vilspa.esa.es/sas/7.0.0/watchout/\-Evergreen\_tips\_and\_tricks/UVflux.html}
As we are only interested in comparing the UV and X-ray fluxes, rather than in performing detailed fitting 
of the Spectral Energy Distribution (SED), we did not deem it worthwhile to 
perform a more sophisticated analysis. In table \ref{UV/O} we report the UV fluxes of the sources, 
corrected for extinction, and the UV/X (2--10 keV) fluxes ratio.

There are large differences between the UV/X fluxes ratios of the sources.  There is no apparent relation with
either the  Black hole mass or the Eddington ratio. Given the rather constant values of the
0.5-2/2-10 keV ratio and of the soft power law index, this suggests that the soft X-ray excess and the UV
flux are not directly related each other. In the table, the ratio between the extrapolation to 231 nm
of the baseline, thermal disc, ionized disc reflection and relativistic absorption models to the observed
values are also given. No model reproduces well the UV flux for all sources. The power law (BMS model)
always overpredicts the UV flux. The disc thermal emission strongly underpredict the UV flux for Ark~374 and
PG~0052+251. It is more difficult to judge the reflection and relativistic disc models, as the extrapolation
to the UV must fall short of the data because these models do not include the disc thermal emission, which
must contribute in the UV. Certainly, at least for Fairall~1116 both models fails (and the wind model for H0439-272 as well).

\begin{table*}
\begin{center}
\begin{tabular}{cccccccc}
\textbf{Source} &\textbf{F}$_{UV}$ & \textbf{F$_{UV}$/F$_{X}$} & $R_{BMS}$ & $R_{Diskbb}$ & $R_{Refl}$ & $R_{RelAbs}$ \\
 &  $\times10^{-12}$ erg cm$^{-2}$ s$^{-1}$  & & & & \\
\hline \textbf{H0439-272}  & 5.29 & 0.92  & 11.2 & 0.36 & 0.17 & 1.1 \\
\hline \textbf{Ark 374}  & 18.8 & 5.70 & 4.0 & 0.09 & 0.05 & 0.3 \\
\hline \textbf{Fairall 1116} &  0.90  & 0.17 & 300 & 1.7 & 1.9 & 9.2 \\
\hline \textbf{PG0052+251} &  35.7  & 5.27 & 3.6 & 0.03 & 0.04 & 0.2 \\
\hline
\end{tabular}
\end{center}
\caption{ First column: UV fluxes, corrected for the extinction; second column: UV/X (2--10 keV) fluxes ratio;
third to sixth columns: the ratio between the extrapolated best fit models to 231 nm and the observed
values.}\label{UV/O}
\end{table*}

\section{Discussion and Conclusions}

\begin{figure}
\epsfig{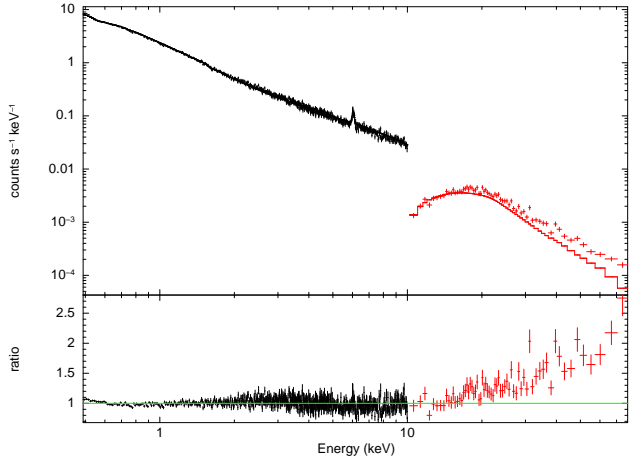}
\caption{A 100 ks Simbol-X simulations of the baseline model spectrum of Fairall 1116 
(upper panel), and the data/model ratio when the spectrum is fitted with the smeared
absorption model (lower panel).}
\label{sx}
\end{figure}

We have analyzed 4 moderately luminous AGN observed with the XMM--$Newton$ satellite. The main results
can be summarized as follows:

a) The hard X-ray emission is characterized by a power law with spectral indices with depends very much
on the modeling of the soft excess. If the latter is modelled as a power law, the hard X-ray photon index
is quite flat, $\Gamma$ ranging from 1.25 to 1.56. If, on the other hand, other parameterization are chosen
(some of them giving fits of comparable quality), much steeper $\Gamma$ are found, similar to those usually
found in AGN (e.g. Porquet et al. 2004, Piconcelli et al. 2005, Bianchi et al. in prep.),
apart when using a smeared absorbing wind, when quite steep values, $\Gamma >$2, are found.

b) An iron line is also detected, even if in PG 0052+251 only
an upper limit is found. The error bars are very large, but the
decrease of the best fit values of the EW with both the luminosity
and the the Eddington ratio is in agreement with what found by
Bianchi et al. (2007) in a much larger sample (the Iwawasa--Taniguchi effect).

c) No satisfactory modeling for the soft excess is found. 

A power law, besides resulting in suspiciously flat hard
power laws, has no obvious physical meaning, and it will not be
discussed further. We just note that the Comptonization model
(see below) is basically a power law (and indeed the fits are 
statistically comparable) with a cutoff (which results in a steeper
hard power law).

A multicolor thermal disc emission, apart from
providing significantly worse fits, results in very high temperatures, almost constant despite
the factor $\sim$10 differences in Black Hole masses, which should result in a factor $\sim$2 differences
in $kT$.

A Comptonization model provides both good fits and ``normal'' hard power law indices, but both
the temperature and the optical depth of the Comptonizing electron are suspiciously
the same for all objects, with no obvious reasons for such a constancy.

A (relativistically blurred) ionized reflection model provides not very good fits (which
may be attributed to the oversimplifications of the model, like a single ionization zone
while a centrally illuminated disc obviously has a strong radial dependence of $\xi$, e.g. 
Matt et al. 1993). Moreover, the range of ionization parameters spanned by the four sources 
is rather small, even if this may be due to the fact that the zone of the disc with that  
ionization parameters are the ones contributing most to the soft excess (smaller ionizations
resulting in a smaller albedo, higher ionizations in a smaller energy dependence of the albedo).

Finally, the relativistically smeared absorbing wind model provides good fits to the data, 
but the three model parameters (column density, ionization parameter and smearing velocity)
are almost constant among the sources. While there may be an explanation for the constancy
of the column and the ionization parameter (see discussion in Middleton et al. 2006), the
constancy of the smearing velocity and, above all, its high value (about 0.5 $c$) are difficult
to explain. Outflowing winds with velocities as high as 0.1-0.2 $c$ have been observed in a 
few high accretion rate sources (e.g. Pounds et al. 2003). Fixing the smearing velocity to
0.15, however, results in unacceptable fits. It is interesting to remark that
Schurch \& Done (2007b), using a more refined version of the model (not availaible
for spectral fitting), found that indeed 
extremely large, unrealistic velocities are required to explain the soft X-ray
excesses in AGN.

In summary, none of the abovementioned model provides a satisfactory description of the data,
even if none of them can be completely ruled out.
The constancy of the parameters  we found for {\it all} models
is clearly telling us that some characteristic energy is involved. Future, high sensitivity
broad band measurements like those provided by e.g. Simbol-X (Ferrando et al. 2006)
will hopefully tell us which, if any,
of these model is tenable (Matt 2007, Ponti et al. 2007). As an example, in Fig.~\ref{sx} the
Simbol-X simulation of the baseline model for Fairall 1116 is shown, as well as the data/model ratio
after fitting with the smeared absorption model.

d) The UV/X ratio is very different from source to source -- more than a factor 30 -- and does
not correlate with any other parameter. It is important to note that the soft--to--hard X-rays
ratio is instead basically constant, suggesting that the UV and soft X--rays are not physically
related -- another clue against a thermal disc origin for the soft X--rays.

\section*{Acknowledgements}

We thank the anonymous referee for valuable comments which helped us 
improving the clarity of the paper.
This paper is based on observations obtained with XMM-Newton, an ESA science
mission with instruments and contributions directly
funded by ESA Member States and the USA (NASA).
GM, SB and EJB acknowledge financial support from ASI under grant I/023/05/0.

\end{document}